# Conical thin shell wormhole from global monopole: A theoretical construction

F.Rahaman*, M.Kalam‡ and K A Rahman*

**Abstract**

By applying 'Darmois-Israel formalism', we establish a new class of thin shell wormhole in the context of global monopole resulting from the breaking of a global O(3) symmetry. Since global monopole is asymptotically conical ( no longer asymptotically flat ), we call it as conical thin shell wormhole. Different characteristics of this conical thin shell wormhole, namely, time evolution of the throat, stability, total amount of exotic matter have been discussed.

# Introduction:

It is believed that during the evolution, the universe had undergone a number of phase transitions. Phase transitions in the early universe can give rise to various forms of topological defects. They can be monopoles, cosmic strings or domain walls[1-3]. Among them monopoles and cosmic strings are well studied for their cosmological as well as astrophysical implications. A global monopole is a heavy object that forms in the phase transition of a system composed of a self coupling scalar field, $\Phi^a$ whose original O(3) symmetry is spontaneously broken to U(1)[4]. There has been a fairly large amount of discussions [5-18] on the gravitational field of global monopoles beginning with the work of Barriola and Vilenkin ( BV )[4]. According to BV, global monopoles have Goldstone fields with energy density decreasing with the distance as $r^{-2}$. BV have also shown that the space time produced by global monopole has no Newtonian gravitational potential in spite of the fact that the geometry produced by this heavy object has a non-vanishing curvature.

The quest for traversable Wormhole in theoretical physics started at the end of 1980's. In two remarkable works, Morris and Thorne[19] and also Morris, Thorne and Yurtsever[20] have shown that these are the solutions of Einstein field equations that have two regions connected by a throat. To get a Wormhole solution, one has to tolerate the violation of null energy condition (NEC). This means a ghost like matter ( i.e. a matter which violates NEC ) distribution should present as a source of Energy Momentum Tensor. Definitely, it is very difficult to deal with this ghost like matter (exotic matter). So, physicists have been trying to minimize the total amount of exotic matter. In a pioneer work, Visser[21] designed a model of minimizing the usage of exotic matter to construct a Wormhole in which the above matter is constructed at the Wormhole throat. This model is constructed by surgically grafting two manifolds to form a geodesically complete region in such a way that no horizon is permitted to form. This theoretical construction of Wormhole is known as thin shell Wormhole. Down to the present moment, various thin shell Wormholes have been discussed. For examples, thin shell Wormholes in Schwarzschild geometry[22], Reissner-Nordström black hole geometries[23], Dilaton fields [24], Einstein-Maxwell theory with a Gauss Bonnet term [25], Higher dimension space time [26], Heterotic string theory [27] and thin shell Wormhole from tidal charged black hole [28].

One of the most challenging current problems in theoretical physics is explaining the structure formation of the Universe. Pando et al [29] have proposed that topological defects are responsible for structure formation of the galaxies. Nucamendi and others [ 30 - 32 ] have suggested that the monopole ( its density being proportional to $\frac{1}{r^2}$ ) could be the galactic dark matter in the spiral galaxies. The gravitational field of global monopole may lead to the clustering in matter and they can induce anisotropies in cosmic microwave background radiation. At this moment there is no contradiction with observation data and there is no observation data which permits rule out definitely the possibility of existence of topological defects [33]. Recently, Eiroa et al [34] and Bejarano et al [35] have studied cylindrical thin shell Wormhole in the context of cosmic strings. Also, several authors have been discussed Wormholes associated cosmic strings [36-38]. Like cosmic string, global monopole shows some peculiar results. It exerts no gravitational force on surrounding matters but space around it has a deficit solid angle and light rays are deflected by the same angle, independent of the impact parameter. So it is of great interest to investigate the different characteristics of the thin shell wormhole constructed by global monopole. We develop the model by cutting and pasting metrics corresponding to BV's monopole. Various aspects of these thin shell Wormhole, namely, temporal, evolution of the throat, stability, total amount of exotic matter will be discussed.

The layout of the paper as follows :

In section 2, the reader is reminded about global monopole solution obtained by BV. In section 3, Thin shell Wormhole has been constructed by means of the cut and paste techniques. In section 4, the time evolution of the radius of the throat is considered whereas linearized stability analysis is studied in section 5. In section 6, the total amount of exotic matter has been calculated. Section 7 is devoted to a brief summary and discussion.

# 2. Global monopole:

In this section, we briefly review the work of BV[4]. The Lagrangian of a global monopole in general relativity has the form

$$L = \frac{1}{2}\partial_\mu \phi^a \partial^\mu \phi^a - \frac{1}{2}\lambda(\phi^a \phi^a - \eta^2)^2 \tag{1}$$

The field configuration describing a monopole is

$$\phi^a = \eta f(r)\frac{x^a}{r} \tag{2}$$

where $x^a x^a = r^2$ ( $a = 1,2,3$ ). The most general static metric admitting spherical symmetry is given by

$$ds^2 = -B(r)dt^2 + A(r)dr^2 + r^2(d\theta^2 + \sin\theta d\phi^2) \tag{3}$$

The field equation for $\phi^a$ reduces to a single equation for f(r):

$$\frac{(r^2 f')'}{Ar^2} + \frac{1}{2B}(\frac{B}{A})' f' - \frac{2f}{r^2} - \lambda\eta^2 f(f^2 - 1) = 0 \tag{4}$$

[ $''$ refers to differentiation with respect to radial coordinate ]
Using Lagrangian (1) and the metric (3), the components of energy momentum tensor can be written via

$$T_{\mu\nu} = 2\frac{\partial L}{\partial g^{\mu\nu}} - L g_{\mu\nu} \tag{5}$$

as follows:

$$T_t^t = -\eta^2[\frac{f^2}{r^2} + \frac{f'^2}{2A} + \frac{\lambda}{4}\eta^2 f(f^2 - 1)] \tag{6}$$

$$T_r^r = -\eta^2[\frac{f^2}{r^2} - \frac{f'^2}{2A} + \frac{\lambda}{4}\eta^2 f(f^2 - 1)] \tag{7}$$

$$T_\theta^\theta = T_\phi^\phi = -\eta^2[\frac{f'^2}{2A} + \frac{\lambda}{4}\eta^2 f(f^2 - 1)] \tag{8}$$

In the flat space the monopole core has a size $\delta \sim \frac{1}{\sqrt{\lambda\eta^2}}$ and mass $M \sim \lambda\eta^4\delta^3 \sim \frac{\eta}{\sqrt{\lambda}}$. Outside the core, one can approximate $f(r) \sim 1$ and equations (6)-(8) simplify to

$$T_t^t = T_r^r = \frac{\eta^2}{r^2}, T_\theta^\theta = T_\phi^\phi = 0 \tag{9}$$

Now using the Einstein's field equations, $R_{\mu\nu} = 8\pi G(T_{\mu\nu} - \frac{1}{2}g_{\mu\nu}T)$, one can obtain the solutions as

$$B = A^{-1} = 1 - 8\pi G\eta^2 - \frac{2GM}{r} \tag{10}$$

where $M$ is a constant of integration and can be considered as the mass of the monopole core.

Thus finally, one gets the metric describing the gravitational field of global monopole as

$$ds^2 = -f(r)dt^2 + \frac{dr^2}{f(r)} + r^2 d\Omega_2^2, \tag{11}$$

where

$$f(r) = 1 - 8\pi G\eta^2 - \frac{2GM}{r} \tag{12}$$

This monopole exerts no gravitational force on non relativistic matter, but the space around it has a deficit solid angle. Neglecting the mass term and rescaling the time and radial coordinates, one can get the monopole metric as

$$ds^2 = -dt^2 + dr^2 + r^2(1 - 8\pi G\eta^2)d\Omega_2^2$$

This metric describes a space with a deficit angle. Actually, the surface $\theta = \frac{\pi}{2}$ has the geometry of a cone with a deficit angle $\Delta = 8\pi^2 G\eta^2$.

## 3. Formation of the conical thin shell wormhole :

In this section, we use Cut and Paste technique to construct conical thin shell wormhole from global monopole spacetimes. Consider two spacetimes $G^+$ and $G^-$, which are endowed with the same metric as ( 11 ) (i.e.corresponding to the same global monopole spacetime). Taking two copies of region from $G^+$ and $G^-$ with $r \geq a$ : $M^{\pm} \equiv (x|r \geq a)$ where $a > \delta$ [ $\delta$ = radius of the monopole core and $\pm$ indicates two copies ], one can paste two copies $M^{\pm}$ together at the hypersurface $\Sigma = \Sigma^{\pm} = (x|r = a)$. This construction creates a geodesically complete manifold $M = M^+ \bigcup M^-$ with two asymtotically conical regions connected by a throat placed at $\Sigma$. Now, one can expect that the surface stresses of this thin junction surface $\Sigma$ are proportional to a delta function. Following Darmois-Israel formalism [39], we shall determine the surface stresses at the junction boundary. The intrinsic coordinates in $\Sigma$ are taken as $\xi^i = (\tau, \theta, \phi)$ with $\tau$ is the proper time on the shell. To understand the dynamics of the wormhole, we assume the radius of the throat be a function of the proper time $a = a(\tau)$. The parametric equation for $\Sigma$ is defined by

$$\Sigma : F(r, \tau) = r - a(\tau) = 0 \tag{13}$$

The second fundamental form ( extrinsic curvature ) associated with the two sides of the shell are

$$K_{ij}^{\pm} = -n_{\nu}^{\pm} \left[\frac{\partial^2 X_\nu}{\partial \xi^i \partial \xi^j} + \Gamma^\nu_{\alpha\beta} \frac{\partial X^\alpha}{\partial \xi^i} \frac{\partial X^\beta}{\partial \xi^j}\right]|_{\Sigma} \tag{14}$$

where $n_\nu^{\pm}$ are the unit normals to $\Sigma$ in M:

$$n_\nu^{\pm} = \pm |g^{\alpha\beta} \frac{\partial F}{\partial X^\alpha} \frac{\partial F}{\partial X^\beta}|^{-\frac{1}{2}} \frac{\partial F}{\partial X^\nu} \tag{15}$$

[ $i, j = 1, 2, 3$ corresponding to boundary $\Sigma$ ; $\alpha, \beta = 1, 2, 3, 4$ corresponding to original spacetime ] with $n^\mu n_\mu = 1$.

The intrinsic metric at $\Sigma$ is given by

$$ds^2 = -d\tau^2 + a(\tau)^2 d\Omega^2 \tag{16}$$

The position of the throat of the wormhole is described by $X^\mu = (t, a(t), \theta, \phi)$. The unit normal to $\Sigma$ is given by

$$n^\nu = (\frac{\dot{a}}{f(a)}, \sqrt{f(a) + \dot{a}^2}, 0, 0) \tag{17}$$

Now using equations (14), (15) and (17), the non trivial components of the extrinsic curvature are given by

$$K^\pm_{\tau\tau} = \mp \frac{\frac{1}{2} f'(a) + \ddot{a}}{\sqrt{f(a) + \dot{a}^2}} \tag{18}$$

$$K^\pm_{\theta\theta} = K^\pm_{\phi\phi} = \pm \frac{1}{a} \sqrt{f(a) + \dot{a}^2} \tag{19}$$

We define jump of the discontinuity of the extrinsic curvature of the two sides of $\Sigma$ as $[K_{ij}] = K^+_{ij} - K^-_{ij}$ and $K = [K^i_i] = trace[K_{ij}]$.

The Ricci tensor at the throat can be calculated in terms of the discontinuity of the second fundamental forms ( extrinsic curvature ). This jump discontinuity, together with Einstein field equations, provides the stress energy tensor of $\Sigma$, where throat is localized: $T^{\mu\nu} = S^{\mu\nu} \delta(\eta)$ [ $\eta$ denotes the proper distance away from the throat ( in the normal direction )] with,

$$S^i_j = -\frac{1}{8\pi}([K^i_j] - \delta^i_j K) \tag{20}$$

where $S^i_j = diag(-\sigma, -v_\theta, -v_\phi)$ is the surface energy tensor with $\sigma$, the surface density and $v_\theta$ and $v_\phi$, the surface tensions. Now taking into account the equation (20), one can find

$$\sigma = -\frac{1}{2\pi a} \sqrt{1 - 8\pi G \eta^2 - \frac{2GM}{a} + \dot{a}^2} \tag{21}$$

$$-v_\theta = -v_\phi = -v = \frac{1}{4\pi a} \frac{1 - 8\pi G \eta^2 - \frac{GM}{a} + \dot{a}^2 + a\ddot{a}}{\sqrt{1 - 8\pi G \eta^2 - \frac{2GM}{a} + \dot{a}^2}} \tag{22}$$

[ over dot and prime mean, respectively, the derivatives with respect to $\tau$ and a ].

Negative surface energy in (21) implies the existence of ghost like matter at the shell. The negative signs of the tensions mean that they are indeed pressures.

# 4. Time evolution of radius of the throat:

Now, we consider static solutions of the shell by replacing $\dot{a} = 0$ and $\ddot{a} = 0$ in equations (21) and (22):

$$\sigma = -\frac{1}{2\pi a}\sqrt{1 - 8\pi G\eta^2 - \frac{2GM}{a}} \tag{23}$$

$$v = -\frac{1}{4\pi a}\frac{1 - 8\pi G\eta^2 - \frac{GM}{a}}{\sqrt{1 - 8\pi G\eta^2 - \frac{2GM}{a}}} \tag{24}$$

Now, one can write the equations (23) and (24) in the form

$$v = w(a)\sigma \tag{25}$$

where

$$w(a) = \frac{1}{2}\frac{1 - 8\pi G\eta^2 - \frac{GM}{a}}{1 - 8\pi G\eta^2 - \frac{2GM}{a}} \tag{26}$$

From equation (26), one can find the equation of state of matter located at the static throat ( i.e. for a given value of the throat radius ). Following Eiroa et al [34], we assume that equation of state does not depend on derivatives of a($\tau$) i.e. form of equation of state kept the same form as in the dynamic case. Now, replacing equations (21) and (22) in (25), we get the following expression as

$$\ddot{a}(1 - 8\pi G\eta^2 - \frac{2GM}{a}) - \dot{a}^2\frac{GM}{a^2} = 0 \tag{27}$$

This implies,

$$\dot{a}(\tau) = \dot{a}(\tau_0)[\frac{1 - 8\pi G\eta^2 - \frac{2GM}{a(\tau)}}{1 - 8\pi G\eta^2 - \frac{2GM}{a(\tau_0)}}]^{\frac{1}{2}} \tag{28}$$

Here, $\tau_0$ is arbitrary fixed time.

Thus one gets,

$$\int_{a(\tau_0)}^{a(\tau)}\frac{1}{\sqrt{1 - 8\pi G\eta^2 - \frac{2GM}{a}}}da = \frac{\dot{a}(\tau_0)(\tau - \tau_0)}{\sqrt{1 - 8\pi G\eta^2 - \frac{2GM}{a(\tau_0)}}} \tag{29}$$

This gives,

$$\frac{\sqrt{(1-8\pi G\eta^2)a^2(\tau) - 2GMa(\tau)}}{(1-8\pi G\eta^2)} + \frac{2GM}{(1-8\pi G\eta^2)^{\frac{3}{2}}}\ln[\sqrt{(1-8\pi G\eta^2)a(\tau)} + \sqrt{(1-8\pi G\eta^2)a(\tau) - 2GM}] = \frac{\dot{a}(\tau_0)(\tau-\tau_0)}{\sqrt{1-8\pi G\eta^2 - \frac{2GM}{a(\tau_0)}}}$$

The above implicit expression gives the time evolution of the radius of the throat.

The velocity and acceleration of the throat are

$$\dot{a}(\tau) = \dot{a}(\tau_0)[\frac{1 - 8\pi G\eta^2 - \frac{2GM}{a(\tau)}}{1 - 8\pi G\eta^2 - \frac{2GM}{a(\tau_0)}}]^{\frac{1}{2}} \tag{30}$$

$$\ddot{a} = \frac{\dot{a}^2(\tau_0)\frac{2GM}{a^2(\tau)}}{(1 - 8\pi G\eta^2 - \frac{2GM}{a(\tau_0)})} \tag{31}$$

From the above two expressions, one can see that the sign of the velocity depends on the sign of the initial velocity but the acceleration is always positive. It is immaterial whether the initial velocity is positive or negative, the throat expands forever. This would imply that the equilibrium position is always unstable. However, if the initial velocity is zero, the velocity and acceleration of the throat would be zero i.e. throat be in static equilibrium position. Now, we shall study the stability of the configuration under small perturbations around static solution situated at $a_0$ ( initial velocity will be assumed to be zero ).

# 5. Linearized Stability Analysis:

Rearranging equation (21), we obtain the thin shell's equation of motion

$$\dot{a}^2 + V(a) = 0 \tag{32}$$

Here the potential is defined as

$$V(a) = 1 - 8\pi G\eta^2 - \frac{2GM}{a} - 4\pi^2 a^2 \sigma^2 \tag{33}$$

# 5.1 Static Solution:

The above single dynamical equation (32) completely determines the motion of the wormhole throat. One can consider a linear perturbation around a static solution with radius $a_0$. We are trying to find a condition for which stress energy tensor components at $a_0$ will obey null energy condition. For a static configuration of radius $a_0$, we obtain respective values of the surface energy density and the surface pressure by using the explicit form of the metric as,

$$\sigma_0 = -\frac{1}{2\pi a_0}\sqrt{1 - 8\pi G\eta^2 - \frac{2GM}{a_0}} \tag{34}$$

Since negative tension is equivalent to pressure, we take,

$$-\upsilon = -\upsilon_\theta = p_\theta = -\upsilon_\phi = p_\phi = p \tag{35}$$

Here,

$$p_0 = \frac{1}{4\pi a_0} \frac{1 - 8\pi G\eta^2 - \frac{GM}{a_0}}{\sqrt{1 - 8\pi G\eta^2 - \frac{2GM}{a_0}}} \tag{36}$$

One can see that surface density is always negative and implying the violation of weak and dominant energy conditions. Now, we check whether null energy condition will be satisfied or violated. Taking the following relationship,

$$\sigma_0 + p_0 = -\frac{(1 - 8\pi G\eta^2 - \frac{3GM}{a_0})}{\sqrt{1 - 8\pi G\eta^2 - \frac{2GM}{a_0}}} \tag{37}$$

The above equation implies $\sigma_0 + p_0 < 0$ i.e. null energy condition is always violated.

Hence the conical thin shell wormhole constructed by joining the geometries corresponding to monopole spacetime, the null energy condition is always violated.

## 5.2 Stability Analysis:

Linearizing around a static solution situated at $a_0$, one can expand V(a) around $a_0$ to yield

$$V = V(a_0) + V'(a_0)(a - a_0) + \frac{1}{2}V''(a_0)(a - a_0)^2 + 0[(a - a_0)^3] \tag{38}$$

where prime denotes derivative with respect to $a$.

Since we are linearizing around a static solution at $a = a_0$, we have $V(a_0) = 0$ and $V'(a_0) = 0$. The stable equilibrium configurations correspond to the condition $V''(a_0) > 0$. Now we define a parameter $\beta$, which is interpreted as the speed of sound, by the relation

$$\beta^2(\sigma) = \frac{\partial p}{\partial \sigma}|_\sigma \tag{39}$$

$$V''(a) = -\frac{4GM}{a^3} - 8\pi^2\sigma^2 - 32\pi^2 a\sigma\sigma' - 8\pi^2 a^2(\sigma')^2 - 8\pi^2 a^2\sigma\sigma'' \tag{40}$$

From equations (21) and (22)( by using (35) ), one can write energy conservation equation as

$$\dot\sigma + 2\frac{\dot a}{a}(p + \sigma) = 0 \tag{41}$$

or

$$\frac{d}{d\tau}(4\pi\sigma a^2) + p\frac{d}{d\tau}(4\pi a^2) = 0 \tag{42}$$

From equation (42) ( by using (40) ), we obtain,

$$\sigma'' + \frac{2}{a}\sigma'(1+\beta^2) - \frac{2}{a^2}(p+\sigma) = 0 \tag{43}$$

The wormhole solution is stable if $V''(a_0) > 0$ i.e. if ( by using (39) and (41) )

$$\beta_0^2 < [\frac{(1-8\pi G\eta^2 - \frac{5GM}{a_0})}{2(1-8\pi G\eta^2 - \frac{3GM}{a_0})} - \frac{(\frac{GM}{a_0})^2}{2(1-8\pi G\eta^2 - \frac{3GM}{a_0})(1-8\pi G\eta^2 - \frac{2GM}{a_0})}] - 1 \tag{44}$$

For monopole configuration, the parameters G, M , $\eta$ are known quantities. So the stability of the configuration requires the above restriction on $\beta_0$. This means there exists some part of the parameter space where the throat location is stable. For a lot of useful information, we show the stability region graphically ( see fig. 1 ).

# 6. Total amount of exotic matter:

We have seen that matter located in the shell violates the weak and null energy conditions. That means $\rho < 0$ and $p_j + \rho < 0 \; \forall$ j, where $\rho = \sigma$, the energy density given in equation (21) and $p_j$, the principal pressures ( here radial pressure, $p_r$ is zero and transverse pressures $p = -v = p_\theta = -v_\theta = p_\phi = -v_\phi$ are given in equation(22)). Now, outside the shell, one can note that $T_t^t = T_r^r = \frac{\eta^2}{r^2}$ and $T_\theta^\theta = T_\phi^\phi = 0$. These mean that $\rho > 0$ as well as $p_j + \rho > 0$, in other words, the weak and null energy conditions are satisfied outside the shell. Therefore, one concludes that the ghost like matter is located only in the shell. Now, we calculate the total amount of ghost like matter for the conical thin shell wormhole.

This can be quantified by the following integrals[40-41]:

$$\Omega_1 = \int \rho\sqrt{-g}d^3x, \Omega_j = \int [\rho + p_j]\sqrt{-g}d^3x \tag{45}$$

Following Eiroa and Simone [24] , we introduce a new radial coordinate $R = \pm(r-a)$ in M ( $\pm$ for $M^\pm$ respectively ) so that

$$\Omega_1 = \int_0^{2\pi}\int_0^\pi\int_{-\infty}^\infty \rho\sqrt{-g}dRd\theta d\phi \tag{46}$$

$$\Omega_j = \int_0^{2\pi}\int_0^\pi\int_{-\infty}^\infty [\rho + p_j]\sqrt{-g}dRd\theta d\phi \tag{47}$$

Since the shell does not exert radial pressure and the energy density is located on a thin shell surface, so that $\rho = \rho + p_r = \delta(R)\sigma_0$, $\rho + p_t = \delta(R)(\sigma_0 + p_0)$,

Hence, one gets,

$$\Omega_r = \Omega_1 = \int_0^{2\pi} \int_0^{\pi} [\sigma\sqrt{-g}]|_{r=a_0} d\theta d\phi = 4\pi a_0^2 \sigma(a_0) = -2a_0 \sqrt{1 - 8\pi G\eta^2 - \frac{2GM}{a_0}} \quad (48)$$

$$\Omega_t = \int_0^{2\pi} \int_0^{\pi} [(\sigma + P)\sqrt{-g}]|_{r=a_0} d\theta d\phi = -a_0 \frac{(1 - 8\pi G\eta^2 - \frac{3GM}{a_0})}{\sqrt{1 - 8\pi G\eta^2 - \frac{2GM}{a_0}}} \quad (49)$$

Thus one could see that the scale of symmetry breaking $\eta$ and mass of the monopole affect the total amount of ghost like matter needed. From the above measures, we note that the total amount of matter would be reduced as desired with the suitable choice of the parameters. The variation of $\Omega_t$ and $\Omega_1$ with respected to scale of symmetry breaking and mass of the monopole are shown in the figures 2 - 5.

# 7. Summary and Discussions:

In this article, we have studied conical thin shell wormhole constructed from global monopole spacetimes. Since, monopoles are the topological defects created at the early stages of the Universe so it may possible to form thin shell like wormhole during phase transitions. We have obtained the time evolution of the radius of throat. One could see that whether the initial velocity is positive or negative, the throat expands indefinitely. That means, this conical thin shell like wormholes are very much unstable. Thats why, it is not seen today. When initial velocity is zero, the radius of the throat remains constant i.e. the throat be a position of static equilibrium. We have analyzed the dynamical stability of the thin shell, considering linearized radial perturbations around static solution. To analyze this, we define a parameter $\beta^2 = \frac{p'}{\sigma'}$ as a parametrization of the stability of equilibrium. We have obtained a restriction on $\beta^2$ to get stable equilibrium of the conical thin shell wormhole( see eq.(44)). We have calculated integral measuring of the total amount of exotic matter. Finally, we have shown that total amount of exotic matter needed to support traversable wormhole can be reduced as desired with the suitable choice of the parameters. The variation of total amount of exotic matters with respect to different parameters are depicted in figures 2-5. From the figures, one can note that total amount of exotic matters will be reduced with the increasing of scale of symmetry breaking as well as with increase of the mass of the monopole.

# Acknowledgments

F.R. is thankful to DST , Government of India for providing financial support. MK has been partially supported by UGC, Government of India under MRP scheme.

**Figure Captions:**

Figure - 1 : We define $p = \frac{M}{a_0}$ and choose $8\pi G\eta^2 \sim 10^{-6}$ [1] and $G = 1$. Here we plot $\beta_{|(a=a_0)}$ $Vs.$ $p$ . The stability region is given below the curve.

Figure - 2 : We choose $y = \frac{\Omega_1}{a_0}$ and $x = \frac{a_0}{M}$. The variation of total amount of exotic matter on the shell with respect to the mass of the monopole is shown in the figure. ( $8\pi G\eta^2 \sim 10^{-6}$ and $G = 1$)

Figure - 3 : We choose $y = \frac{\Omega_1}{a_0}$ and $x = \eta$. The variation of total amount of exotic matter on the shell with respect to the scale of symmetry breaking $\eta$ is shown in the figure. ( solid line for $\frac{2M}{a_0} = .01$ and dotted line for $\frac{2M}{a_0} = .02$)

Figure - 4 : We choose $y = \frac{\Omega_t}{a_0}$ and $x = \frac{a_0}{M}$. The variation of total amount of exotic matter on the shell with respect to the mass of the monopole is shown in the figure. ( $8\pi G\eta^2 \sim 10^{-6}$ and $G = 1$)

Figure - 5 : We choose $y = \frac{\Omega_t}{a_0}$ and $x = \eta$. The variation of total amount of exotic matter on the shell with respect to the scale of symmetry breaking $\eta$ is shown in the figure. ( solid line for $\frac{M}{a_0} = .01$ and dotted line for $\frac{M}{a_0} = .03$)

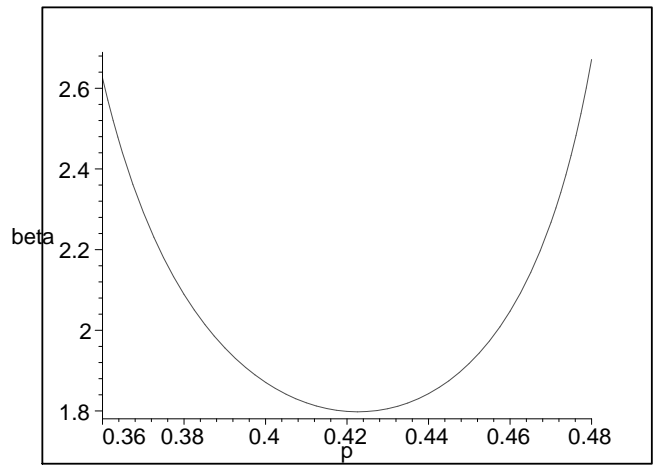

Figure 1:

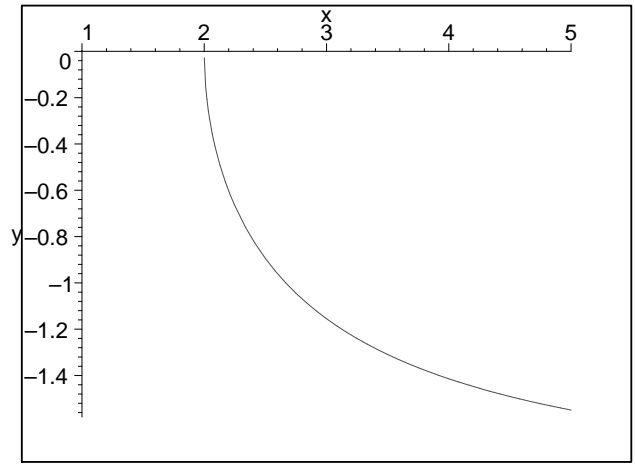

Figure 2:

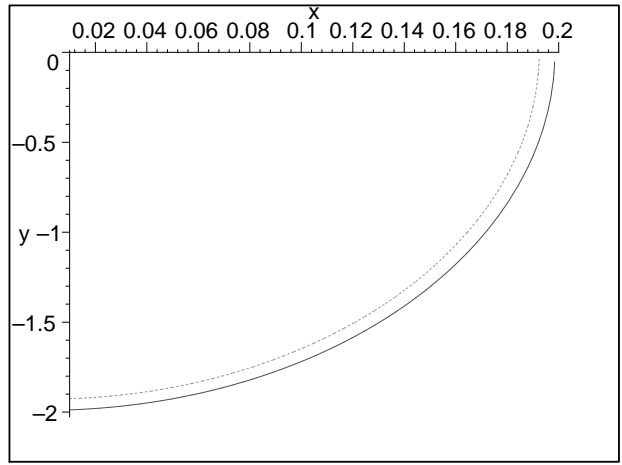

Figure 3:

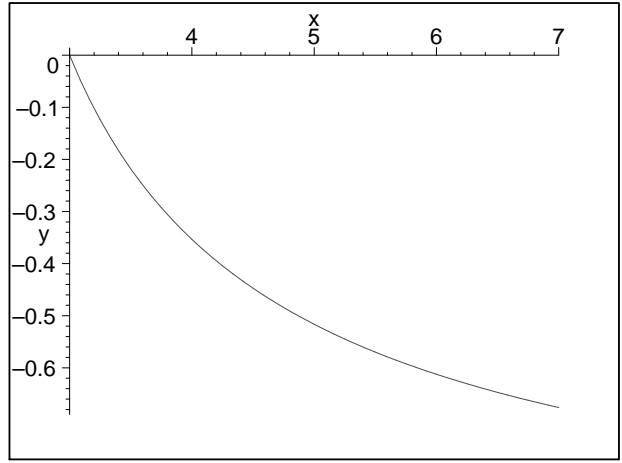

Figure 4:

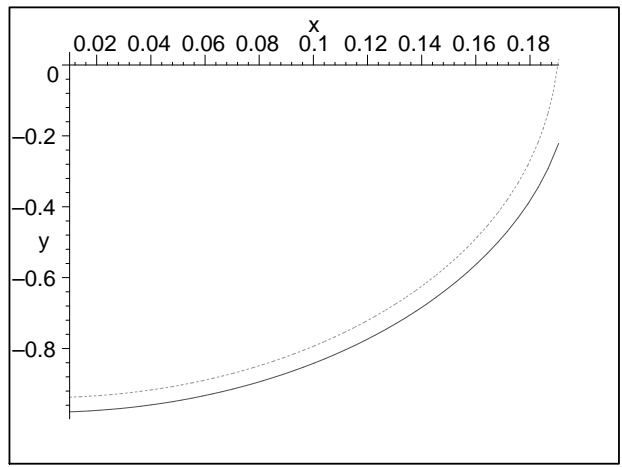

Figure 5:

15